\documentclass[iop]{emulateapj}

\usepackage{graphics,epsf,graphicx,longtable,natbib,amsmath}
\usepackage[T1]{fontenc}
\usepackage{mathptmx}

\newcommand{\kms}{\ifmmode{~{\rm km\,s}^{-1}}\else{~km~s$^{-1}$~}\fi}
\newcommand{\msun}{\ifmmode{{\rm M}_\odot}\else{${\rm M}_\odot$}\fi}

\defcitealias{bode06}{Paper I}
\defcitealias{vaytet07}{Paper II}

\slugcomment{}

\shorttitle{X-RAY SPECTRAL MODELLING OF RS OPH (2006)}
\shortauthors{VAYTET ET AL.}

\begin{document}

\title{{\it Swift} observations of the 2006 outburst of the recurrent nova RS Ophiuchi:\\
       III. X-ray spectral modelling}

\author{N.~M.~H. Vaytet\footnotemark[1,2], T.~J. O'Brien\footnotemark[3], K.~L. Page\footnotemark[4], M.~F. Bode\footnotemark[5], M. Lloyd\footnotemark[6] \textsc{and} A.~P. Beardmore\footnotemark[7]}

\footnotetext[1]{CEA/DSM/IRFU, Service d'Astrophysique, Laboratoire AIM, CNRS, Universit\'{e} Paris Diderot, 91191 Gif-sur-Yvette, Cedex, France; neil.vaytet@cea.fr}

\footnotetext[2]{Ecole Normale Sup\'{e}rieure de Lyon, CRAL (UMR CNRS 5574), 69364 Lyon Cedex 07, France; neil.vaytet@ens-lyon.fr}

\footnotetext[3]{Jodrell Bank Centre for Astrophysics, School of Physics and Astronomy, The University of Manchester, Manchester, M13 9PL, UK; tim.obrien@manchester.ac.uk}

\footnotetext[4]{Department of Physics and Astronomy, University of Leicester, Leicester, LE1 7RH, UK; kpa@star.le.ac.uk}

\footnotetext[5]{Astrophysics Research Institute, Liverpool John Moores University, Birkenhead, CH41 1LD, UK; mfb@astro.livjm.ac.uk}

\footnotetext[6]{Jodrell Bank Centre for Astrophysics, School of Physics and Astronomy, The University of Manchester, Manchester, M13 9PL, UK; myfanwy.lloyd@manchester.ac.uk}

\footnotetext[7]{Department of Physics and Astronomy, University of Leicester, Leicester, LE1 7RH, UK; apb@star.le.ac.uk}

\begin{abstract}
Following the {\it Swift} X-ray observations of the 2006 outburst of the recurrent nova RS Ophiuchi, we developed hydrodynamical models of mass ejection from which the forward shock velocities were used to estimate the ejecta mass and velocity. In order to further constrain our model parameters, here we present synthetic X-ray spectra from our hydrodynamical calculations which we compare to the {\it Swift} data. An extensive set of simulations was carried out to find a model which best fits the spectra up to 100 days after outburst. We find a good fit at high energies but require additional absorption to match the low energy emission. We estimate the ejecta mass to be in the range $(2-5) \times 10^{-7}~\msun$ and the ejection velocity to be greater than $6000\kms$ (and probably closer to $10,\!000\kms$). We also find that estimates of shock velocity derived from gas temperatures via standard model fits to the X-ray spectra are much lower than the true shock velocities.
\end{abstract}


\section{Introduction}\label{sec:intro}

RS Ophiuchi (RS Oph) is one of the most well-studied of the ten known recurrent novae (RNe; \citealt{anupama08}). The central system is a binary comprising a white dwarf (WD) and a red giant (RG) \citep{dobrzycka94}. A classical nova outburst involves mass transfer from the companion to the WD. The build-up of pressure and temperature in the degenerate layer of accreted hydrogen eventually leads to a thermonuclear runaway (TNR) on the surface of the WD, resulting in the high-speed ejection of a shell of material into the circumstellar medium \citep{starrfield08}. Novae which have been observed in outburst more than once are called RNe. The shock interaction of the ejecta with the surrounding medium has been found to heat the gas to temperatures of $10^{7} - 10^{8}$ K, yielding hard X-ray radiation (\citealt{lloyd92}; \citealt*{obrien94}; \citealt{mukai01}). In the case of RS Oph, the ejecta run into the surrounding dense RG wind. Soft X-ray emission is also expected to be revealed later in the outburst from a central WD close to Eddington luminosity (\citealt{krautter96}; \citealt{balman98}) as is seen in RS Oph (see below).

RS Oph has had outbursts recorded in 1898, 1933, 1958, 1967, 1985 (see \citealt{rosino87a}; \citealt{rosino87b}) and most recently on 2006 February 12 \citep{narumi06}, with possible additional outbursts in 1907 and 1945 (\citealt{schaefer04}; \citealt{oppenheimer93}). Its binary system has an orbital period of $455.72\pm 0.83$ days \citep{fekel00}. The mass of the WD is suspected to be close to the Chandrasekhar limit from various considerations, including the speed class of outburst and the relatively short recurrence time between outbursts (e.g. \citealt{yaron05}, \citealt{sokoloski06}, \citealt{hachisu07}). The distance to RS Oph is 1.6 kpc (\citealt{bode87}; see also \citealt{barry08}).

RS Oph was detected in X-rays just 2 days after the 2006 outburst with the RXTE satellite \citep{sokoloski06} and 3.17 days after outburst using the XRT instrument on board the {\it Swift} satellite (\citealt{bode06}, hereafter \citetalias{bode06}). The {\it Swift} observations were performed on average every one or two days during the first 100 days, covering all the phases of the remnant's evolution. The rate of observations subsequently decreased but continued for more than a year \citep{page08}. RS Oph was seen initially as a source of fairly hard X-rays, gradually softening with time. Around 26 days after outburst, the very bright and soft component from the Super-Soft Source (SSS) phase of nuclear burning on the surface of the WD appeared in the spectrum (\citealt{osborne06,hachisu07,page08,osborne11}).

In an effort to improve on the models of the 1985 outburst and, in particular, to address the new observations of the early phases of its evolution, we developed new hydrodynamical models for RS Oph (\citealt*{vaytet07}, hereafter \citetalias{vaytet07}). We used a one-dimensional Eulerian second order Godunov code \citep{obrien94} to simulate the outburst where mass-loss in the form of a fast wind from the WD ran into a surrounding slow RG wind. The scheme took into account all three phases of the remnant's evolution, and allowed us, after a thorough analysis of the shock velocities, to determine that the ejecta had a low mass ($\sim2\times 10^{-7}~\msun$) and a high ejection velocity. The models also showed that radiative cooling had an important impact on the evolution of the shocks; the amount of energy radiated by the hot ejecta was correlated to the rate at which the forward shock velocity decreased in time.
 
In \citetalias{vaytet07} our analysis relied on single-temperature fits to the {\it Swift} X-ray spectra (from \citetalias{bode06}) in order to determine post-shock temperatures and hence the shock velocities. Here we improve on this approach by calculating full X-ray spectra directly from our hydrodynamical results using \textsc{xspec}\footnote{http://swift.gsfc.nasa.gov/docs/software/lheasoft} and by including the effects of overlying absorption and instrument response. Results from an extensive set of simulations are compared to the {\it Swift} observations for the first 100 days after outburst in order to obtain best-fitting values for the model parameters. 

\section{{\it Swift} XRT observations}\label{sec:obs}

Some 70 observational epochs of RS Ophiuchi were obtained using the X-ray Telescope (XRT) onboard the {\it Swift} observatory during the first 100 days after the 2006 outburst. As in \citetalias{bode06}, the data were processed using the standard {\it Swift} software and source spectra were extracted from the cleaned Windowed Timing (WT) mode ($t < 90$ days) and Photon Counting (PC) mode ($t > 90$ days) event lists. Because of the high count rate of the X-ray source, most of the WT were piled-up. Thus, the core of the PSF was excluded: an outer box of width $60\times20$ pixels was used, with the central 10 pixels (1 pixel = $2\arcsec36$) removed. Similarly, the PC data were piled-up: here, an annulus with outer radius 30 pixels was used, with the inner exclusion radius decreasing from six to zero pixels between days 90 and 100. Background spectra were obtained from nearby source-free regions, using a $60\times20$ pixel box for the WT data and a 60 pixel radius circle for the PC. We corrected for the fractional loss of the point-spread function caused both by the exclusion of the central (piled-up) region and the positioning of the source over the bad CCD columns (caused by micrometeorite damage on 2005 May 2007; \citealt{abbey06}) when calculating the count rate and flux.

Grades 0--2 and 0--12 were chosen for the spectral analysis of the WT and PC data respectively. The FTOOL \texttt{xrtmkarf} was used to generate suitable ancillary response function files, and these were used in conjunction with the relevant response matrices. All spectra were grouped to a minimum of 20 counts per bin in order to facilitate the use of $\chi^{2}$ statistics in \textsc{xspec}, the results of which are shown in Fig.~\ref{fig:spectra} (black crosses) for 8 epochs.

\section{Calculation of synthetic X-ray spectra}\label{sec:calcxrayspec}

Our hydrodynamical models from \citetalias{vaytet07} predict the physical conditions of the hot shocked gas at any time during the outburst. In practise the model comprises many concentric spherical shells each with a different temperature and density. Since it would be too computationally intensive to calculate the X-ray emission from each shell separately, we grouped them into 30 logarithmically-divided temperature bins from $10^4$ to $10^9$~K, weighting each shell's contribution by its emission measure (experiment suggested 30 bins were sufficient to give an accurate result). These were then fed as a 30-component plasma into \textsc{xspec} (version 12), an X-ray data reduction and spectrum calculation code. This was used to compute the X-ray spectrum with the MEKAL model for emission from hot, diffuse gas, including line emissions from C, N, O, Ne, Na, Mg, Al, Si, S, Ar, Ca, Fe and Ni, along with free-free, free-bound and two-photon emission (for more details see \citealt{vaytet09}; \citealt{mewe86}).

The interstellar equivalent hydrogen absorption column for our models was taken from \citet{hjellming86} who found $n_{H} = (2.4 \pm 0.6) \times 10^{21} ~\mathrm{cm}^{-2}$. We have also included a contribution from the circumstellar absorption due to the RG wind surrounding the hot, shocked gas behind the forward shock. At each epoch we calculate an average value for this circumstellar absorption across the face of the sphere of shocked gas. 

Of course, only the unshocked neutral part of the RG wind will contribute to absorption. It is however non-trivial to know the ionisation state of the RG wind at all times, since a large portion of it could be ionised by the radiation emitted during the outburst. This requires not only knowledge of the recombination rates in the RG wind but also knowledge of the luminosity of the central radiation source at all times, which is clearly not constant nor simple to model (see for example \citealt{page08}). \citet*{obrien92} assumed that the whole RG wind was flash-ionised by UV radiation at the very beginning of the outburst, then computed recombination rates inside the RG wind. Here we simply use upper and lower extremes; either the RG wind is fully neutral or fully ionised. We are then confident that the correct absorption has to lie between these limits since they do not depend on any assumptions about the central radiation field.

The final step in the creation of the synthetic spectra was to fold the results through the {\it Swift} XRT instrumental response matrix. The response used here was the same as for the observations mentioned above, with the corresponding ancillary file specific to each observation to provide the precise effective area of the detector and correct for bad channels. We used the canonical distance of 1.6 kpc for normalisation. 

\section{Results}\label{sec:results}

\subsection{Initial simulations from \citetalias{vaytet07}}\label{sec:initialresults}

Model runs 1 to 16 presented in this section are identical to those in \citetalias{vaytet07}; see Table~\ref{tab:params} for a list of parameter values. Run 14 is not included since it represented a continuous fast wind ejection during the entire duration of the simulation, thus rendering parameters such as ejected mass $M_{\mathrm{ej}}$ and outburst energy $E_{0}$ meaningless.

\begin{figure*}[!ht]
\begin{center}
\includegraphics[scale=0.42]{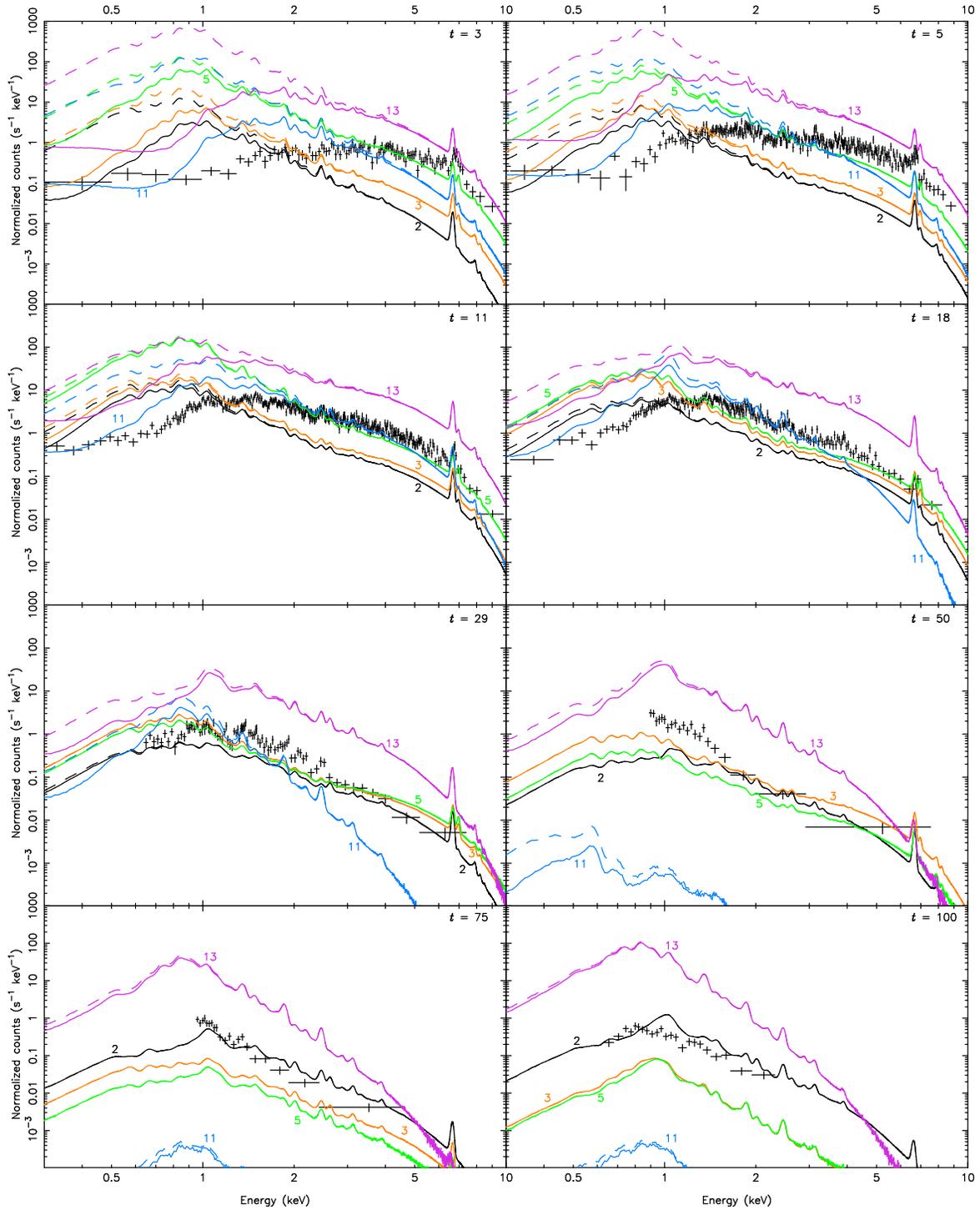}
\caption[Initial X-ray spectra]{Synthetic X-ray spectra for runs 2 (black), 3 (orange), 5 (green), 11 (blue) and 13 (dark pink) at days 3, 5, 11, 18, 29, 50, 75 and 100. The solid and dashed lines represent a fully neutral and fully ionised RG wind, respectively. The black crosses represent the {\it Swift} data. The {\it Swift} data at low energies for days 29 and after showing the SSS component have been removed since no SSS is included in our models.}
\label{fig:spectra}
\end{center}
\end{figure*}

Fig.~\ref{fig:spectra} shows the synthetic spectra for five of these simulations compared with the {\it Swift} observations. Results are shown in separate panels for days 3, 5, 11, 18, 29, 50, 75 and 100 after outburst. Each panel includes results assuming the overlying RG wind is either fully neutral (solid line) or fully ionised (dashed line). We first consider the results from run 2 (shown in black), a fairly typical example. At early times ($3-18$ days), before the SSS component from the WD remnant appears at the soft end of the spectrum, the synthetic spectra clearly do not match the observations; the model is overestimating the soft X-ray count rates and underestimating the hard count rates. We do not include the SSS component in our models so for later times (day 29 and after), the {\it Swift} data at the soft end of the spectrum has been omitted (energies $\lesssim 1$ keV). At these late times, at the hard end of the spectrum the model count rates are very closely matched to the observations, suggesting that the hard count rates have decreased faster in the observations than in the model.

In order to link the synthetic spectra with the hydrodynamics, we recall Fig.~5 in \citetalias{vaytet07} where we concluded that the deceleration rate of the forward shock was an important aspect of the shock models. We concluded that runs 3 and 11 have forward shock deceleration rates lower and higher than run 2, respectively. Their X-ray spectra are shown in Fig.~\ref{fig:spectra} (orange for run 3 and blue for run 11). 

In run 3 (orange), the increased $M_{\mathrm{ej}}$ increases the X-ray count rates, this effect being strongest at the hard end of the spectrum. The hard X-ray count rates decrease more slowly than in run 2, which is due to the higher inertia that the shell carries due to its higher mass. This is best seen at days 18 to 50. Days 75 and 100 should be ignored in this case as by this time the shock has reached the edge of the RG wind; see Fig.~4 in \citetalias{vaytet07}. 

In the case of run 11 (blue), the high fast wind velocity $V_{2}$ also increases the X-ray count rates, but this time over the whole energy spectrum, most probably simply due to the higher outburst energy compared to run 2. In run 11 both hard and soft X-ray count rates decrease much faster than in runs 2 and 3, falling to the bottom of the visible plot area by day 75. Note that for run 11 the shock does not reach the edge of the wind on a 100-day timescale. Finally, the impact of absorption in the overlying high-density neutral RG wind in run 11 (solid blue line) is clearly visible on the soft part of the spectrum at early times where the predicted count rates are more consistent with observations.

Further, by simply increasing $V_{2}$ and leaving $M_{\mathrm{ej}}$ untouched as in run 5 (Fig.~\ref{fig:spectra}; green lines), we note that the count rates are also greatly increased across the spectrum. Runs 3 and 5 have identical outburst energies and RG wind densities; it is hence not surprising to see that their X-ray spectra are very similar at late times (day 29 and after) apart from the hard count rates being slightly higher in run 3 at day 50. Comparing run 5 with run 11 reveals that increasing $V_{2}$ can only be effective in enhancing hard count rates at early times if $M_{\mathrm{ej}}$ is not too low.

We conclude that, as one might expect, greater deceleration of the forward shock results in faster reduction in hard X-ray count rates. This is also sensitive to the ejected mass $M_{\mathrm{ej}}$; a high-mass shell will carry more momentum and therefore decelerate more slowly. A high ejection velocity $(V_{2} \ge 5000\kms)$ greatly overestimates the soft count rates. Note that the strong Fe XXV K line at 6.7 keV appears wider in the {\it Swift} observations than in our models (run 2 for example). We do not include Doppler broadening in our models but it should also be noted that a second Fe XXVI line (6.97 keV) appears very close to the first in some of our models (runs 3, 5 and 13) which would not be resolved by the XRT and would broaden the appearance of the Fe line. Finally, it is clear from runs 2 and 3 that a relatively low-density (even entirely neutral) RG wind appears not to produce sufficient absorption of low-energy X-rays at early times.

In \citetalias{vaytet07}, run 13 provided the best fit to the forward shock velocities. However the extremely high implied $V_{2}$ of 16,300~km~s$^{-1}$, for which there was little evidence elsewhere, cast some doubt on the validity of the model. The synthetic spectra for run 13 are shown in Fig.~\ref{fig:spectra} (dark pink). These clearly show run 13 does not match the observations of the 2006 outburst of RS Oph. Although the desired large amount of absorption is present at early times, the count rates are overestimated across the spectrum at every epoch.

\subsection{Fitting synthetic spectra to the {\it Swift} data}\label{sec:spectrafitting}

A fuller exploration of the parameter space of the hydrodynamical model has now been undertaken by carrying out a new extensive set of simulations using the same code as in \citetalias{vaytet07}. This brings the total number of runs to 88. All the parameters for runs 1 to 88 are listed in Table~\ref{tab:params} (run 14 excluded for reasons given above).

\onecolumngrid
\begin{center}
\begin{longtable}{@{~}c@{~}c@{~}c@{~}c@{~}c@{~}c@{~}c@{~}c@{~}c@{~}c@{~}c@{~}c@{~}c@{~}}
\caption{\textsc{ASPHERE code parameter space}}
\label{tab:params}
\\
\hline
\hline

   \multicolumn{1}{@{~}c@{~}}{Run}        & \multicolumn{1}{@{~}c@{~}}{RG wind}  & \multicolumn{1}{@{~}c@{~}}{Fast wind} & 
   \multicolumn{1}{@{~}c@{~}}{Fast wind}  & \multicolumn{1}{@{~}c@{~}}{Outburst} & \multicolumn{1}{@{~}c@{~}}{Ejected}   & \multicolumn{1}{@{~}c@{~}}{RG wind} & 
   \multicolumn{1}{@{~}c@{~}}{Mass ratio} & \multicolumn{1}{@{~}c@{~}}{RG wind}  & \multicolumn{1}{@{~}c@{~}}{Fast wind} & \multicolumn{3}{@{~}c@{~}}{Mean flux difference}\\
   \cline{11-13}
   
   \multicolumn{1}{@{~}c@{~}}{$~$}                                & \multicolumn{1}{@{~}c@{~}}{mass-loss} & \multicolumn{1}{@{~}c@{~}}{mass-loss} & 
   \multicolumn{1}{@{~}c@{~}}{velocity}                           & \multicolumn{1}{@{~}c@{~}}{energy}    & \multicolumn{1}{@{~}c@{~}}{mass}      & \multicolumn{1}{@{~}c@{~}}{mass}  & 
   \multicolumn{1}{@{~}c@{~}}{$M_{\mathrm{ej}}/M_{\mathrm{RGW}}$} & \multicolumn{1}{@{~}c@{~}}{ratio}     & \multicolumn{1}{@{~}c@{~}}{phase}     & \multicolumn{1}{@{~}c@{~}}{total} & \multicolumn{1}{@{~}c@{~}}{early} & \multicolumn{1}{@{~}c@{~}}{late} \\
   
   \multicolumn{1}{@{~}c@{~}}{$~$}     & \multicolumn{1}{@{~}c@{~}}{$\dot{M_{1}}$}   & \multicolumn{1}{@{~}c@{~}}{$\dot{M_{2}}$}     & 
   \multicolumn{1}{@{~}c@{~}}{$V_{2}$} & \multicolumn{1}{@{~}c@{~}}{$E_{0}$}         & \multicolumn{1}{@{~}c@{~}}{$M_{\mathrm{ej}}$} & \multicolumn{1}{@{~}c@{~}}{$M_{\mathrm{RGW}}$} & 
   \multicolumn{1}{@{~}c@{~}}{$R$}     & \multicolumn{1}{@{~}c@{~}}{$\dot{M_{1}}/u$} & \multicolumn{1}{@{~}c@{~}}{$t_{0}$}           & \multicolumn{1}{@{~}c@{~}}{$\Phi_{T}$}         & \multicolumn{1}{@{~}c@{~}}{$\Phi_{E}$} & \multicolumn{1}{@{~}c@{~}}{$\Phi_{L}$} \\
   
   \multicolumn{1}{@{~}c@{~}}{$~$}        & \multicolumn{1}{@{~}c@{~}}{($\!\times 10^{-7}$ \msun/yr)}  & \multicolumn{1}{@{~}c@{~}}{($\!\times 10^{-7}$ \msun/yr)}   & 
   \multicolumn{1}{@{~}c@{~}}{($\!\kms$)} & \multicolumn{1}{@{~}c@{~}}{($\!\times 10^{43}$ erg)}       & \multicolumn{1}{@{~}c@{~}}{($\!\times 10^{-6}~\msun$)}      & \multicolumn{1}{@{~}c@{~}}{($\!\times 10^{-6}~\msun$)} & 
   \multicolumn{1}{@{~}c@{~}}{$~$}        & \multicolumn{1}{@{~}c@{~}}{($\!\times 10^{12}$ g/cm)}      & \multicolumn{1}{@{~}c@{~}}{(days)}                          & \multicolumn{1}{@{~}c@{~}}{$~$}                        & \multicolumn{1}{@{~}c@{~}}{($t < 30$d)} & \multicolumn{1}{@{~}c@{~}}{($t \geq 30$d)} \\
   \hline
\endfirsthead

\multicolumn{13}{@{~}c@{~}}{{\tablename} \thetable{} -- Continued} \\
   \hline \hline
   \multicolumn{1}{@{~}c@{~}}{Run}     & \multicolumn{1}{@{~}c@{~}}{$\dot{M_{1}}$}   & \multicolumn{1}{@{~}c@{~}}{$\dot{M_{2}}$}     &
   \multicolumn{1}{@{~}c@{~}}{$V_{2}$} & \multicolumn{1}{@{~}c@{~}}{$E_{0}$}         & \multicolumn{1}{@{~}c@{~}}{$M_{\mathrm{ej}}$} & \multicolumn{1}{@{~}c@{~}}{$M_{\mathrm{RGW}}$} & 
   \multicolumn{1}{@{~}c@{~}}{$R$}     & \multicolumn{1}{@{~}c@{~}}{$\dot{M_{1}}/u$} & \multicolumn{1}{@{~}c@{~}}{$t_{0}$}           & \multicolumn{1}{@{~}c@{~}}{$\Phi_{T}$}         & \multicolumn{1}{@{~}c@{~}}{$\Phi_{E}$} & \multicolumn{1}{@{~}c@{~}}{$\Phi_{L}$} \\ \hline
\endhead

   \hline
\endfoot
   
   \hline \hline
   ~\\
   \multicolumn{13}{@{~}c@{~}}{\textsc{Notes.}--The slow wind velocity is $u = 15\kms$ throughout (\citetalias{vaytet07}). Run 14 is omitted (see section~\ref{sec:initialresults} and \citetalias{vaytet07}).}
\endlastfoot

 1 &  1.43 &  158.59 &  3000 &  1.85 & 0.24 &   3.00 & 0.080 &   6.00 &  7.0 &  0.961 & 0.940 &  0.965\\
 2 &  1.43 &  730.62 &  3000 &  8.62 & 1.10 &   3.00 & 0.367 &   6.00 &  7.0 &  0.574 & 0.748 &  0.537\\
 3 &  1.43 & 3171.80 &  3000 & 36.94 & 4.78 &   3.00 & 1.595 &   6.00 &  7.0 &  0.484 & 0.570 &  0.465\\
 4 &  1.43 &  730.62 &  1398 &  1.85 & 1.10 &   3.00 & 0.367 &   6.00 &  7.0 &  0.997 & 0.995 &  0.998\\
 5 &  1.43 &  730.62 &  6251 & 36.94 & 1.10 &   3.00 & 0.367 &   6.00 &  7.0 &  0.611 & 0.311 &  0.676\\
 6 &  1.43 &  730.62 &  3500 & 11.58 & 1.10 &   3.00 & 0.367 &   6.00 &  7.0 &  0.442 & 0.661 &  0.395\\
 7 &  1.43 &  730.62 &  4000 & 15.13 & 1.10 &   3.00 & 0.367 &   6.00 &  7.0 &  0.411 & 0.567 &  0.378\\
 8 &  1.43 & 5113.73 &  3000 &  8.62 & 1.10 &   3.00 & 0.367 &   6.00 &  1.0 &  0.528 & 0.628 &  0.506\\
 9 &  1.43 &  340.97 &  3000 &  8.62 & 1.10 &   3.00 & 0.367 &   6.00 & 15.0 &  0.660 & 0.746 &  0.642\\
10 &  6.34 &  730.62 &  3000 &  8.62 & 1.10 &  13.32 & 0.083 &  26.67 &  7.0 &  0.995 & 0.974 &  1.000\\
11 &  6.34 &   79.30 & 12000 & 14.77 & 0.12 &  13.32 & 0.009 &  26.67 &  7.0 &  0.853 & 0.259 &  0.980\\
12 &  1.43 & 1148.03 &  3969 &  8.62 & 1.10 &   3.00 & 0.367 &   6.00 &  7.0 &  0.567 & 0.697 &  0.539\\
13 &  9.52 &  158.59 & 16300 & 46.73 & 0.20 &  19.98 & 0.010 &  40.00 &  6.0 & 17.956 & 6.491 & 20.413\\
15 &  1.43 &  298.62 &  3000 &  3.48 & 0.45 &   3.00 & 0.150 &   6.00 &  7.0 &  0.634 & 0.849 &  0.588\\
16 &  1.43 &  497.66 &  3000 &  5.80 & 0.75 &   3.00 & 0.250 &   6.00 &  7.0 &  0.620 & 0.786 &  0.584\\
17 &  7.93 &  158.59 &  9000 &  9.50 & 0.14 &  16.65 & 0.008 &  33.33 &  4.0 &  0.986 & 0.921 &  1.000\\
18 & 52.33 &  852.29 &  4500 & 19.14 & 1.10 & 110.00 & 0.010 & 220.20 &  6.0 &  4.160 & 0.728 &  4.895\\
19 &  7.93 &  774.82 &  8000 & 55.00 & 1.00 &  16.65 & 0.060 &  33.33 &  6.0 & 19.847 & 6.348 & 22.740\\
20 &  6.34 &  126.87 & 10000 & 16.42 & 0.19 &  13.32 & 0.014 &  26.67 &  7.0 &  0.748 & 0.375 &  0.828\\
21 &  6.34 &  174.45 &  7000 & 11.06 & 0.26 &  13.32 & 0.020 &  26.67 &  7.0 &  0.975 & 0.863 &  1.000\\
22 &  7.93 &  451.57 &  7000 & 24.54 & 0.58 &  16.65 & 0.035 &  33.33 &  6.0 &  1.099 & 1.276 &  1.061\\
23 &  7.93 &  323.49 &  8000 & 22.96 & 0.42 &  16.65 & 0.025 &  33.33 &  6.0 &  0.908 & 0.877 &  0.915\\
24 &  7.93 &  646.97 &  7000 & 35.16 & 0.83 &  16.65 & 0.050 &  33.33 &  6.0 & 10.241 & 3.533 & 11.678\\
25 &  7.93 &  646.97 &  6000 & 25.83 & 0.83 &  16.65 & 0.050 &  33.33 &  6.0 &  1.223 & 1.534 &  1.157\\
26 &  7.93 &  646.97 &  5000 & 17.94 & 0.83 &  16.65 & 0.050 &  33.33 &  6.0 &  0.939 & 0.655 &  0.999\\
27 &  7.93 &  774.82 &  4500 & 17.40 & 1.00 &  16.65 & 0.060 &  33.33 &  6.0 &  0.958 & 0.763 &  0.999\\
28 &  7.93 & 1035.15 &  4500 & 23.25 & 1.34 &  16.65 & 0.080 &  33.33 &  6.0 &  0.775 & 0.408 &  0.853\\
29 &  7.93 & 1293.95 &  4500 & 29.06 & 1.67 &  16.65 & 0.100 &  33.33 &  6.0 &  2.973 & 1.578 &  3.272\\
30 &  7.93 & 1164.56 &  4000 & 20.67 & 1.50 &  16.65 & 0.090 &  33.33 &  6.0 &  0.901 & 0.466 &  0.995\\
31 &  7.93 & 1578.30 &  4500 & 35.45 & 2.04 &  16.65 & 0.122 &  33.33 &  6.0 &  8.992 & 2.471 & 10.390\\
32 &  7.93 & 1940.92 &  4500 & 43.59 & 2.51 &  16.65 & 0.150 &  33.33 &  6.0 & 14.352 & 3.447 & 16.688\\
33 &  7.93 & 2587.89 &  4500 & 58.12 & 3.34 &  16.65 & 0.201 &  33.33 &  6.0 & 17.011 & 4.900 & 19.606\\
34 &  7.93 & 1578.30 &  4000 & 28.01 & 2.04 &  16.65 & 0.122 &  33.33 &  6.0 &  2.124 & 1.351 &  2.289\\
35 &  7.93 & 1940.92 &  4000 & 34.44 & 2.51 &  16.65 & 0.150 &  33.33 &  6.0 &  8.035 & 2.174 &  9.291\\
36 &  7.93 & 2587.89 &  4000 & 45.92 & 3.34 &  16.65 & 0.201 &  33.33 &  6.0 & 14.954 & 3.296 & 17.452\\
37 &  7.93 & 1578.30 &  3500 & 21.44 & 2.04 &  16.65 & 0.122 &  33.33 &  6.0 &  0.862 & 0.368 &  0.968\\
38 &  7.93 & 1940.92 &  3500 & 26.37 & 2.51 &  16.65 & 0.150 &  33.33 &  6.0 &  0.986 & 0.915 &  1.001\\
39 &  7.93 & 2587.89 &  3500 & 35.16 & 3.34 &  16.65 & 0.201 &  33.33 &  6.0 &  7.962 & 2.065 &  9.225\\
40 &  7.93 &  155.28 & 12000 & 24.80 & 0.20 &  16.65 & 0.012 &  33.33 &  6.0 &  1.366 & 1.745 &  1.284\\
41 &  7.93 &  182.39 & 12000 & 29.13 & 0.24 &  16.65 & 0.014 &  33.33 &  6.0 &  4.023 & 2.979 &  4.247\\
42 &  7.93 &  232.91 & 12000 & 37.20 & 0.30 &  16.65 & 0.018 &  33.33 &  6.0 & 13.453 & 4.468 & 15.378\\
43 &  7.93 &  284.67 & 12000 & 45.46 & 0.37 &  16.65 & 0.022 &  33.33 &  6.0 & 19.062 & 5.674 & 21.931\\
44 &  7.93 &  155.28 & 10500 & 18.99 & 0.20 &  16.65 & 0.012 &  33.33 &  6.0 &  0.885 & 0.349 &  0.999\\
45 &  7.93 &  182.39 & 10500 & 22.30 & 0.24 &  16.65 & 0.014 &  33.33 &  6.0 &  0.845 & 0.750 &  0.866\\
46 &  7.93 &  232.91 & 10500 & 28.48 & 0.30 &  16.65 & 0.018 &  33.33 &  6.0 &  3.182 & 2.736 &  3.277\\
47 &  7.93 &  284.67 & 10500 & 34.81 & 0.37 &  16.65 & 0.022 &  33.33 &  6.0 & 10.973 & 3.949 & 12.478\\
48 &  7.93 &  155.28 &  9300 & 14.89 & 0.20 &  16.65 & 0.012 &  33.33 &  6.0 &  0.969 & 0.827 &  1.000\\
49 &  7.93 &  182.39 &  9300 & 17.50 & 0.24 &  16.65 & 0.014 &  33.33 &  6.0 &  0.938 & 0.649 &  0.999\\
50 &  7.93 &  232.91 &  9300 & 22.34 & 0.30 &  16.65 & 0.018 &  33.33 &  6.0 &  0.841 & 0.730 &  0.865\\
51 &  7.93 &  284.67 &  9300 & 27.31 & 0.37 &  16.65 & 0.022 &  33.33 &  6.0 &  2.154 & 2.374 &  2.107\\
52 &  7.93 &  905.76 &  7000 & 49.22 & 1.17 &  16.65 & 0.070 &  33.33 &  6.0 & 19.007 & 5.449 & 21.912\\
53 &  7.93 & 1100.23 &  7000 & 59.79 & 1.42 &  16.65 & 0.085 &  33.33 &  6.0 & 19.088 & 6.562 & 21.773\\
54 &  7.93 & 1397.76 &  7000 & 75.96 & 1.80 &  16.65 & 0.108 &  33.33 &  6.0 & 16.828 & 7.970 & 18.727\\
55 &  7.93 &  905.76 &  6300 & 39.87 & 1.17 &  16.65 & 0.070 &  33.33 &  6.0 & 13.575 & 4.101 & 15.606\\
56 &  7.93 & 1100.23 &  6300 & 48.43 & 1.42 &  16.65 & 0.085 &  33.33 &  6.0 & 18.193 & 5.128 & 20.993\\
57 &  7.93 & 1397.76 &  6300 & 61.53 & 1.80 &  16.65 & 0.108 &  33.33 &  6.0 & 18.542 & 6.502 & 21.122\\
58 &  7.93 &  905.76 &  5500 & 30.39 & 1.17 &  16.65 & 0.070 &  33.33 &  6.0 &  4.534 & 2.305 &  5.011\\
59 &  7.93 & 1100.23 &  5500 & 36.91 & 1.42 &  16.65 & 0.085 &  33.33 &  6.0 & 10.771 & 3.322 & 12.368\\
60 &  7.93 & 1397.76 &  5500 & 46.89 & 1.80 &  16.65 & 0.108 &  33.33 &  6.0 & 16.799 & 4.490 & 19.437\\
61 &  1.90 & 1278.39 &  4000 & 15.12 & 1.10 &   4.00 & 0.275 &   8.00 &  4.0 &  0.446 & 0.381 &  0.460\\
62 &  1.90 & 1110.13 &  4000 & 13.13 & 0.96 &   4.00 & 0.239 &   8.00 &  4.0 &  0.549 & 0.406 &  0.579\\
63 &  1.90 &  951.54 &  4000 & 11.26 & 0.82 &   4.00 & 0.205 &   8.00 &  4.0 &  0.649 & 0.426 &  0.697\\
64 &  1.90 &  792.95 &  4000 &  9.38 & 0.68 &   4.00 & 0.171 &   8.00 &  4.0 &  0.761 & 0.473 &  0.823\\
65 &  1.90 &  634.36 &  4000 &  7.50 & 0.55 &   4.00 & 0.137 &   8.00 &  4.0 &  0.713 & 0.529 &  0.753\\
66 &  1.90 &  475.77 &  4000 &  5.63 & 0.41 &   4.00 & 0.102 &   8.00 &  4.0 &  0.553 & 0.604 &  0.542\\
67 &  1.90 &  622.94 &  4000 &  5.53 & 0.40 &   4.00 & 0.101 &   8.00 &  3.0 &  0.561 & 0.585 &  0.555\\
68 &  1.90 &  817.64 &  4000 &  4.84 & 0.35 &   4.00 & 0.088 &   8.00 &  2.0 &  0.508 & 0.582 &  0.492\\
69 &  1.90 &  560.68 &  4000 &  4.15 & 0.30 &   4.00 & 0.075 &   8.00 &  2.5 &  0.481 & 0.638 &  0.448\\
70 &  1.90 &  507.49 &  4500 &  5.70 & 0.33 &   4.00 & 0.082 &   8.00 &  3.0 &  0.567 & 0.542 &  0.573\\
71 &  1.90 &  364.76 &  4500 &  5.46 & 0.32 &   4.00 & 0.079 &   8.00 &  4.0 &  0.522 & 0.577 &  0.511\\
72 &  1.90 &  293.39 &  5000 &  6.78 & 0.32 &   4.00 & 0.079 &   8.00 &  5.0 &  0.635 & 0.505 &  0.663\\
73 &  1.90 &  285.46 &  5500 &  7.98 & 0.31 &   4.00 & 0.077 &   8.00 &  5.0 &  0.744 & 0.427 &  0.812\\
74 &  1.90 &  285.46 &  6000 &  9.50 & 0.31 &   4.00 & 0.077 &   8.00 &  5.0 &  0.760 & 0.389 &  0.840\\
75 &  1.90 &  269.60 &  6500 & 10.53 & 0.29 &   4.00 & 0.073 &   8.00 &  5.0 &  0.702 & 0.448 &  0.757\\
76 &  1.43 &  198.24 &  8000 & 11.73 & 0.21 &   3.00 & 0.071 &   6.00 &  5.0 &  0.366 & 0.340 &  0.372\\
77 &  4.76 &  713.66 &  5000 & 16.49 & 0.77 &  10.00 & 0.077 &  20.00 &  5.0 &  3.179 & 0.732 &  3.703\\
78 &  1.90 &  182.38 &  8000 & 15.10 & 0.28 &   4.00 & 0.069 &   8.00 &  7.0 &  0.378 & 0.300 &  0.394\\
79 &  7.93 & 1744.49 &  4500 & 45.71 & 1.13 &  16.65 & 0.068 &  33.33 &  3.0 &  0.859 & 0.274 &  0.984\\
80 &  7.93 & 2061.67 &  5000 & 28.58 & 1.33 &  16.65 & 0.080 &  33.33 &  3.0 &  5.522 & 2.841 &  6.097\\
81 &  1.90 &  158.59 & 10000 & 23.45 & 0.27 &   4.00 & 0.068 &   8.00 &  8.0 &  0.462 & 0.657 &  0.420\\
82 &  1.90 &  111.01 & 10000 & 20.52 & 0.24 &   4.00 & 0.060 &   8.00 & 10.0 &  0.462 & 0.637 &  0.424\\
83 &  1.90 &  364.76 & 10000 & 20.23 & 0.24 &   4.00 & 0.059 &   8.00 &  3.0 &  0.461 & 0.762 &  0.397\\
84 &  1.90 &  491.63 &  6000 &  9.81 & 0.32 &   4.00 & 0.079 &   8.00 &  3.0 &  0.749 & 0.474 &  0.807\\
85 &  1.90 &  388.55 &  8000 & 10.16 & 0.32 &   4.00 & 0.080 &   8.00 &  6.0 &  0.747 & 0.367 &  0.828\\
86 &  1.90 &  888.10 &  6000 & 11.82 & 0.38 &   4.00 & 0.096 &   8.00 &  2.0 &  0.590 & 0.442 &  0.621\\
87 &  7.93 &  951.54 &  5000 & 26.38 & 1.23 &  16.65 & 0.074 &  33.33 &  6.0 &  1.132 & 1.289 &  1.098\\
88 &  1.43 &  730.62 &  8000 & 60.45 & 1.10 &   3.00 & 0.367 &   6.00 &  7.0 &  0.514 & 0.660 &  0.495
\end{longtable}
\end{center}
\twocolumngrid

In \citetalias{vaytet07}, we concluded that the deceleration rate of the forward shock was directly related to the amount of energy radiated away by the ejecta. This in turn appeared to be related to the ratio of the ejected mass to the mass contained in the RG wind; the lower the ratio, the higher the fraction of energy which was radiated away. In Table.~\ref{tab:params} we also list this ratio $R = M_{\mathrm{ej}}/M_{\mathrm{RGW}}$  for each run.

In order to assess the goodness of fit of each run's synthetic spectra to the {\it Swift} data, we extracted fluxes from our models in order to compare them directly to the fluxes observed in some 70 {\it Swift} observations between day 3 and day 100. In all cases, the fluxes have been taken in the energy range $1-10$ keV in order to avoid contamination from the SSS emission. The absolute value of the difference between the model and observed fluxes as a fraction of the observed flux was calculated for each epoch. These were then summed over all epochs and an average was taken, giving the total mean fractional flux difference $\Phi_{T}$. We also define two other quantites $\Phi_{E}$ and $\Phi_{L}$ which are the early ($t < 30$ days) and late ($30 < t < 100$ days) mean fractional flux differences respectively. These values are listed for all runs in Table~\ref{tab:params}. All the values quoted are for runs with a fully-neutral RG wind. It is evident from the {\it Swift} observations that some absorption is present in the spectra, and listing the mean flux differences for the synthetic spectra with an ionised RG wind is therefore unecessary. Moreover, the effect of circumstellar absorption is highest at the soft end of the energy spectrum (see Fig.~\ref{fig:spectra}) and since we only consider energies in the range $1-10$ keV, total mean flux differences for ionised and neutral RG winds are in almost all cases very similar.

The total mean flux differences are a useful way of identifying the regions in the parameter space which yield best fitting runs, however it is not an absolute measure of the quality of the fits. For instance run 11 would be classed as one of the best fitting runs according to $\Phi_{E}$ (Table~\ref{tab:params}), but the model spectra at days 3 and 5 (see Fig.~\ref{fig:spectra}) are clearly not good fits to the data as the spectral slope is very different to the observations. The over-estimate in medium energy counts ($1.0-2.5$ keV) compensates for the underestimate at the hard energy end, giving a total flux very close to the observed flux. We exclude any runs with large $\Phi$ values. We also examined all the spectra by eye to ensure that we had not excluded runs which had the correct spectral shape but were lying either above or below the {\it Swift} data at all epochs. Indeed, such a systematic error could be corrected simply by varying the distance to RS Oph, which has been subject to some debate \citep{barry08}. After this exhaustive analysis, we have come to the following conclusions (ejecta masses are given in $10^{-6}~\msun$):
\begin{itemize}
\item For all runs, in order to reproduce the correct amount of hard X-ray emission, we require $V_{2} \ge 6000\kms$
\item If $M_{\mathrm{ej}} > 1.5 ~ (\times 10^{-6}~\msun)$, spectra can be a good match at early times ($t < 5$ days) but are much too bright at later times (e.g. run 33)
\item If $0.8 < M_{\mathrm{ej}} \le 1.5$:
   \subitem - if $\dot{M_{1}}$ is high, the X-ray flux is either too high (run 19) or fades too quickly (run 25)
   \subitem - if $\dot{M_{1}}$ is low, a high $V_{2} \ge 6000\kms$ forces the fluxes to be too high and the deceleration rate of the forward shock to be too low (run 88)
\item If $0.2 < M_{\mathrm{ej}} \le 0.8$:
   \subitem - if $\dot{M_{1}}$ is high, the X-ray flux is either too low for very low masses (run 48) or fades too quickly (runs 23, 45)
   \subitem - if $\dot{M_{1}}$ is low, hard X-ray counts are too low for $V_{2} < 8000\kms$ (run 73)
   \subitem - a good fit is found for the hard X-rays at all epochs if $\dot{M_{1}}$ is low and $V_{2} \ge 8000\kms$
\item If $M_{\mathrm{ej}} < 0.2$, the hard X-ray flux is too low and the fluxes at all energies fade very rapidly (runs 11, 17)
\end{itemize}

\begin{figure*}[!ht]
\begin{center}
\includegraphics[scale=0.42]{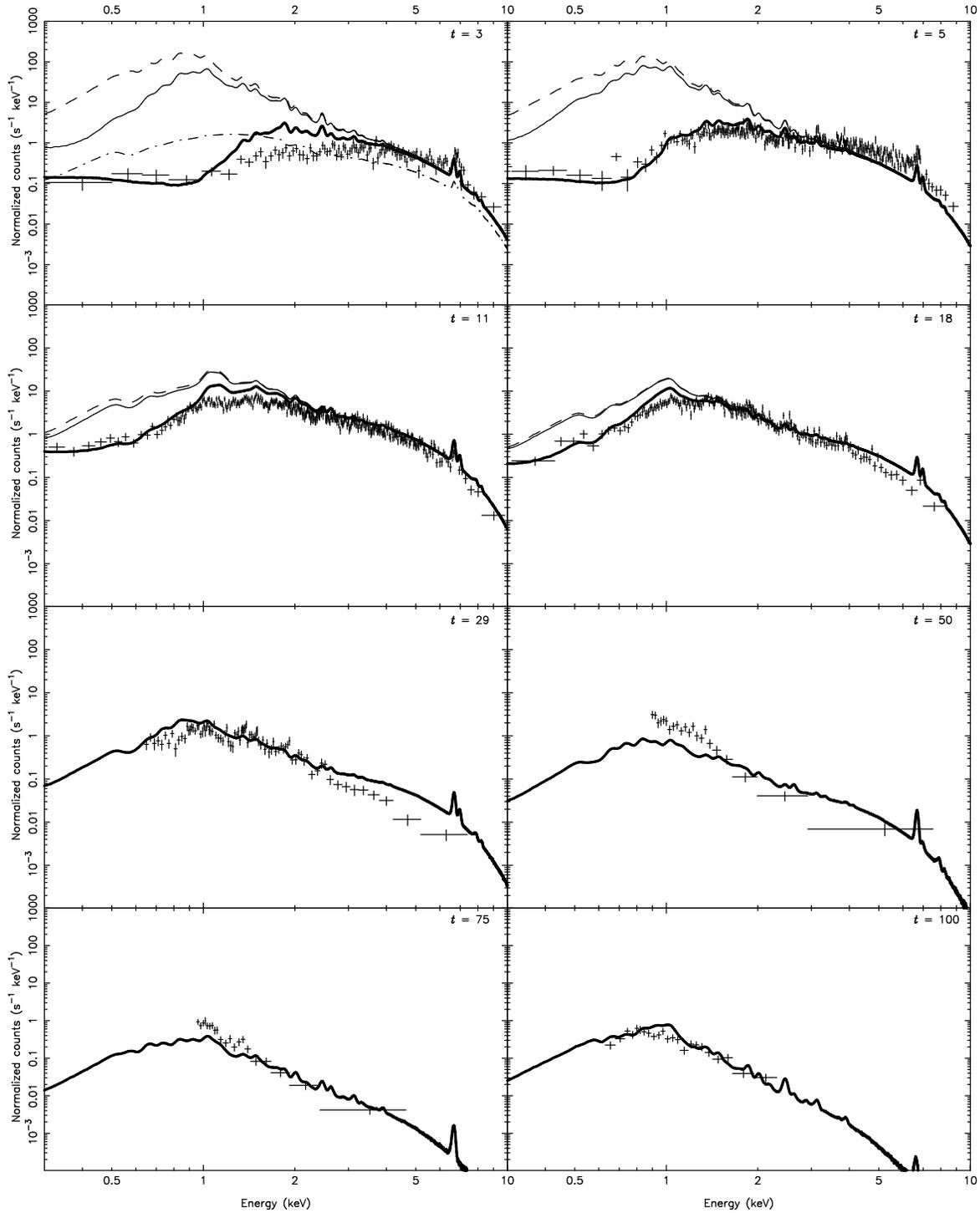}
\caption[Run 81 X-ray spectra]{Synthetic X-ray spectra for run 81 at days 3, 5, 11, 18, 29, 50, 75 and 100. The thin solid and dashed lines represent a fully neutral and fully ionised RG wind, respectively. The bold solid line represents the spectrum including additional absorption. The additional hydrogen equivalent absorption columns are (in units of $10^{-21} ~\mathrm{cm}^{-2}$): 4.0 (day 3), 2.5 (day 5), 0.5 (day 11) and 0.3 (day 18). The black crosses represent the {\it Swift} data. The dot-dash line in the top left panel represents the contribution to the total X-ray spectra from the post-forward shock material (see end of section~\ref{sec:velandflux}).}
\label{fig:spec81}
\end{center}
\end{figure*}

This step by step elimination process leads us to believe that the mass ejected in the outburst $M_{\mathrm{ej}}$ is in the range $(2-8)\times 10^{-7}~\msun$. We find our best fitting model to be run 81 (see Fig.~\ref{fig:spec81}), although even this does show a small overestimate in hard X-rays around day 29. Its main characteristics are its low $M_{\mathrm{ej}} = 2.7 \times 10^{-7}~\msun$ and high $V_{2} = 10^{4} \kms$.

The majority of the runs produce too many low energy X-rays. We have investigated whether this might be due to insufficient circumstellar absorption. In order to naturally increase the absorption, we simply increased the slow wind mass-loss rate (see for example the difference in the amount of absorption in runs 2 and 11) but of course this has a significant effect on the shock dynamics. With a denser RG wind, if we wish to produce shock velocities high enough to be consistent with observations, we have to vastly increase $V_{2}$. This leads to an overestimate of the soft and medium energy count rates ~(see run 13 for example). In order to obtain flat spectra at early times (days 3 and 5, say), we have had to keep $V_{2}$ in the range $\sim 5000-10000\kms$ and consequently the mass-loss rate into the RG wind $\dot{M}_{1}$ also needs to be relatively low. A mid-range $M_{\mathrm{ej}}$ was then required to obtain the correct hard X-ray decay rate, as in run 81.

An alternative way of addressing the deficit in absorption of soft X-rays is to include some additional ad-hoc absorption during the calculation of the spectrum (shown as bold solid lines in Fig.~\ref{fig:spec81}). This allows us to produce good agreement between the synthetic spectra and the {\it Swift} observations. The resulting total absorption column (including the ISM, the CSM from the RG wind and the ad-hoc additional absorption) is within the same order of magnitude as the absorption column required by the fits to the {\it Swift} data in \citetalias{bode06}. Such additional absorption could be due to the presence of a dense gas torus close to the binary core, as suggested by the IR observations of \citet{evans07}.

Of course, we have to acknowledge that such a dense torus would have an effect on the dynamics of the shocks propagating through the CSM but this is a problem that can only be addressed in 2D or 3D simulations. Furthermore, such a dense torus will play a role in the bipolarity seen in the radio \citep{obrien06, sokoloski08, rupen08} and optical imaging \citep{bode07, ribeiro09}. Indeed, \citet{walder08} have shown in full 3D hydrodynamical simulations that the orbiting accreting WD leaves a spiral wake in the RG wind along with a slight equatorial density enhancement, and that the high-density accretion disk around the WD has a substantial size ($3\times10^{12}$ cm in diameter). They illustrate the ejecta collimation efficiency of these features during the nova outburst. \citet{orlando09} have also computed the X-ray emission from 3D simulations of the outburst in a wind with an equatorial density enhancement but unfortunately do not show the absorption effect of the presence of such an enhancement on the resulting synthetic spectra.

\subsection{Shock position, velocities and flux comparisons}\label{sec:velandflux}

\begin{figure*}[!ht]
\begin{center}
\includegraphics[scale=0.60]{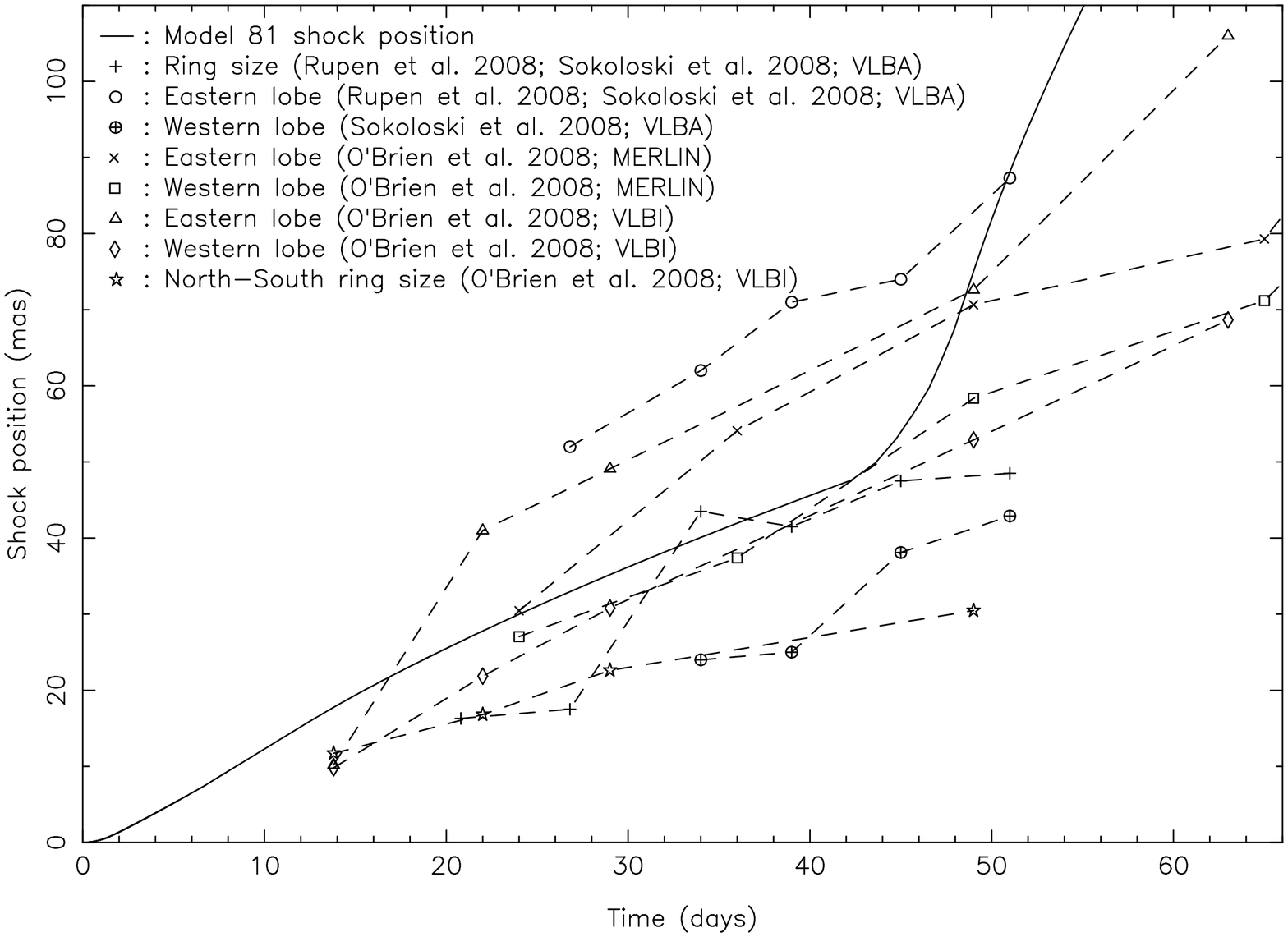}
\caption[Run 81 shock positions]{Run 81 forward shock position as a function of time (solid line) compared to the radio imaging measurements of the size of the expanding ring and the position of the eastern and western lobes by \citet{obrien08}, \citet{rupen08} and \citet{sokoloski08}. The error bars have been omitted for clarity; on the VLBI and VLBA measurements they are $\sim 1$ mas, on the MERLIN data they are $\sim 10$ mas.}
\label{fig:shockpos}
\end{center}
\end{figure*}

Figure~\ref{fig:shockpos} shows the run 81 forward shock position in milli-arcseconds (mas) assuming a distance of 1.6 kpc as a function of time (black solid line), compared to the size measurements of three separate features in resolved radio-interferometry images. The radio images of \citet{obrien08}, \citet{rupen08} and \citet{sokoloski08} have indeed revealed three main synchrotron emitting features (most probably associated with shocks) in the remnant of RS Oph: a central expanding ring, an eastern lobe further away from the central binary system and an additional western counterpart. The positions of the lobes and the size of the expanding ring measured in the radio observations are represented by the various points (see legend for details).

It seems natural to compare the position of our spherically symmetric shock to the size of the expanding ring. We first note that the position of the run 81 forward shock is higher than the early ring size measurements ($t < 30$ days) by a factor of about two. At later times ($30 < t < 50$ days), it is very consistent with the ring size measurements of \citet{sokoloski08}, before it breaks out of the RG wind (kink circa day 44). Between days 20 and 45, the model shock position is also close to the position of the western lobe measured by \citet{obrien08}. It is however never close to the measured positions of the eastern lobe until the shock breakout. We conclude that our model shock positions are consistent with a distance of 1.6 kpc and that over-estimates at early times arise from the fact that we are trying to reproduce the X-ray emission of a system with multiple ejection features moving at different speeds with a spherically symmetric model. The model shock positions sensibly lie between the ring sizes and the eastern lobe positions, as a compromise between the two.

\begin{figure*}[!ht]
\begin{center}
\includegraphics[scale=0.65]{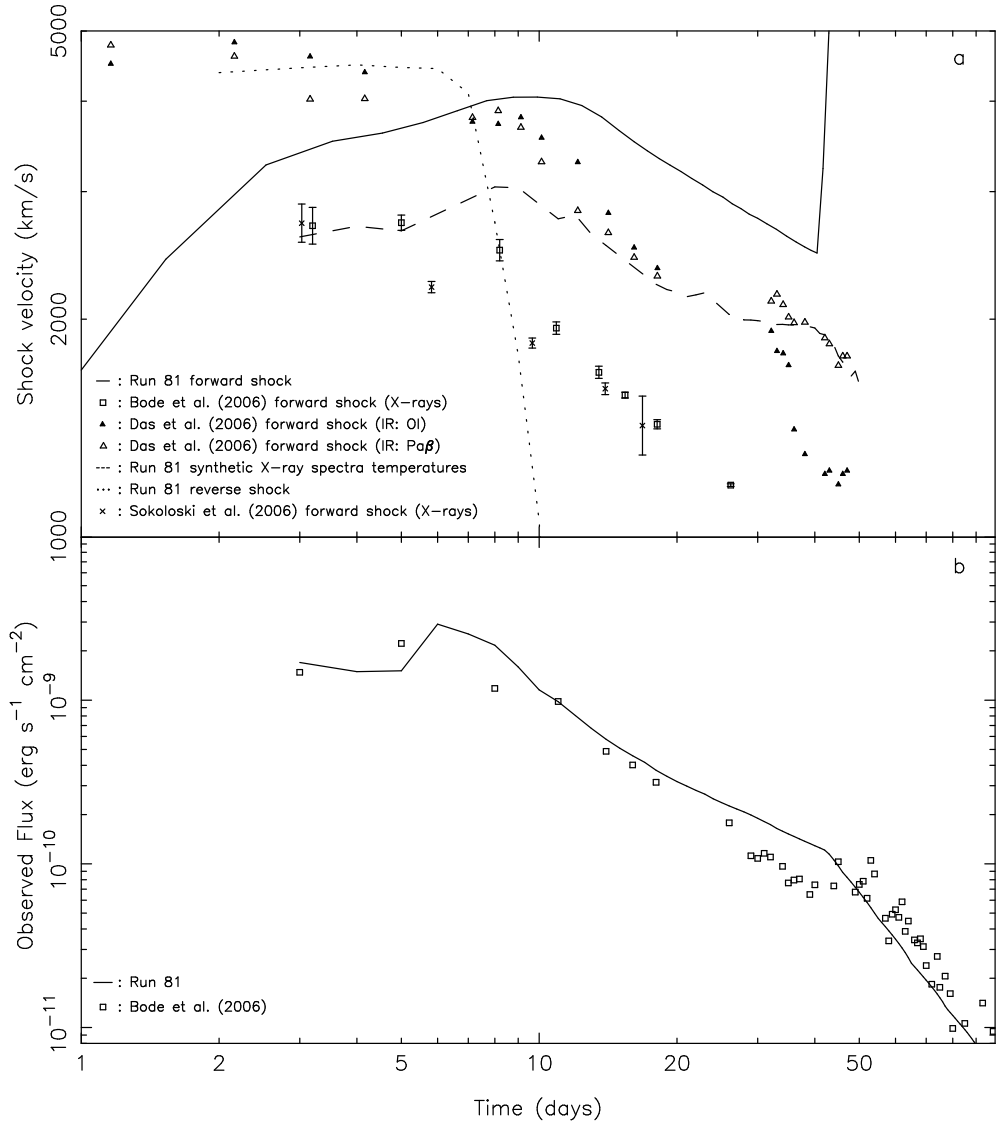}
\caption[Run 81 fluxes and shock velocities]{Run 81 results compared to {\it Swift} data from \citetalias{bode06} and IR data from \citet{das06}. (a) Forward shock velocities: the black empty squares represent the {\it Swift} data, the red triangles the IR data (filled = O{\sc i}; empty = Pa$\beta$) and the solid line the model. The IR velocities were taken from the FWZI/2 of the emission lines. The dashed line is the velocities derived from the $T$ fits to our synthetic spectra. The dotted line represents the gas velocity behind the reverse shock. The crosses are the \citet{sokoloski06} RXTE data. (b) X-ray flux (in the range $1-10$ keV): the empty squares represent the {\it Swift} observations and the solid line the model. Error bars here are smaller than the size of the squares ($<5\%$).}
\label{fig:shockvel}
\end{center}
\end{figure*}

Figure~\ref{fig:shockvel}a shows the forward shock velocities (solid line) measured directly from the hydrodynamical simulation of run 81. These are compared to the shock velocities estimated in \citetalias{bode06} from single-temperature fits to the {\it Swift} spectra (empty squares with error bars) and to the IR velocities from \citet*{das06} (triangles). The IR velocities were taken from the full width zero intensity (FWZI) of the O{\sc i} and Pa$\beta$ emission lines. The velocities from the RXTE X-ray data of \citet{sokoloski06} are also shown (crosses).

Figure~\ref{fig:shockvel}b shows the total $1-10$ keV X-ray fluxes from the {\it Swift} observations (empty squares) compared to the fluxes calculated from run 81 (solid line; including the additional ad-hoc absorption). The model fluxes are consistent with the observed fluxes. The early turnover after an initial rise around day 6 occurs at the same time in both model and observations. There is also a clear break in the model curve just after day 40 which appears to fit the data; this corresponds to the time when the model forward shock breaks out of the RG wind. This occurs at the same time as the dramatic increase in the model velocities.

While the X-ray fluxes are well reproduced, the model velocities do not agree with the \citetalias{bode06} velocities derived from the X-ray observations. The model velocities are much higher, the early turn-over occurs about $4-5$ days later and the deceleration rate is slower. There is however reasonable agreement between the velocities obtained in \citetalias{bode06} and those by \citet{sokoloski06} from RXTE observations. It has been pointed out by \citet{tatischeff08} and references therein that equation (3) in \citetalias{bode06} for strong shock dynamics can underestimate shock velocities where non-linear diffusive shock acceleration of particles is efficient, which is most probably the case for RS Oph. We also note that the immediately post-shock gas is the hottest and would provide the best estimate of the shock velocity. However, the observed spectrum is made up from contributions from a wide range of gas temperatures, emission from the shocked ejecta and even perhaps the WD remnant. These will lead to a softening of the spectrum and so velocities computed from single-temperature fits will be biased towards lower temperatures and hence lower derived shock velocities.

The IR velocities shown in Fig.~\ref{fig:shockvel}a are in better agreement with the model forward shock velocities. However it should be emphasised that the different methods are likely to be biased towards emission from different parts of the expanding remnant. This spans a wide range of velocities, including the unshocked and shocked fast wind from the WD, and the shocked (and unshocked) RG wind. We also show in Fig.~\ref{fig:shockvel}a the gas velocity just after the reverse shock (blue solid line), which is in very good agreement with the IR velocities at early times. Taking the FWZI of the IR lines probes only the fastest moving material in the system, which suggests that the IR line widths are first dominated by the post reverse shock gas and only later by the post forward shock gas.

As previously discussed, in this spherically symmetric model the shock wave breaks out of the RG wind just after day 40, leading to a dramatic increase in model forward shock velocities. By contrast, around the same time (just after day 30), a decrease in the IR O{\sc i} velocities is observed. However, our models show that the accelerating breakout material is at very low density and also very hot so it will most probably not be visible in the IR. After shock breakout, the IR emission comes from other regions of the shell where the velocity is much lower, thus explaining the strong decrease in IR velocities.

In order to investigate the differences between the directly measured shock velocities in our simulations and the velocities estimated from the {\it Swift} X-ray spectra, we fitted two-temperature MEKAL models to our synthetic spectra for run 81. Single-temperature fits were initially attempted but were badly constrained at the high energy end and inconsistent with the synthetic spectra. We then estimated forward shock velocities using the higher of the two temperatures and these are also plotted in Fig.~\ref{fig:shockvel}a (green solid line). These new velocities are in very good agreement with the with the velocities derived by \citetalias{bode06} from {\it Swift} observations at early times ($t < 5$ days) and are much lower than the shock velocities found in the hydrodynamic simulation. They then evolve in a similar way to the hydrodynamic velocities until around day 20, when they start to decrease more slowly than the simulation and the velocities derived from {\it Swift} observations. Note that the hard X-ray count rates in our synthetic model between days 18 and 50 are slightly greater than the {\it Swift} count rates (see Fig.~\ref{fig:spec81}); this yields higher temperatures in the fits and consequently higher velocities. This explains why the slope of the velocity curve derived from our new spectral fits is lower than the slope observed in the {\it Swift} data during that period. At the shock break-out, the computed velocities from the synthetic spectral fits show a sharp decrease, much like the IR velocities, in contrast to the hydrodynamic velocities which increase dramatically, which can be explained in the same way as for the IR velocities (see above).

\begin{figure*}[!ht]
\begin{center}
\includegraphics[scale=0.35]{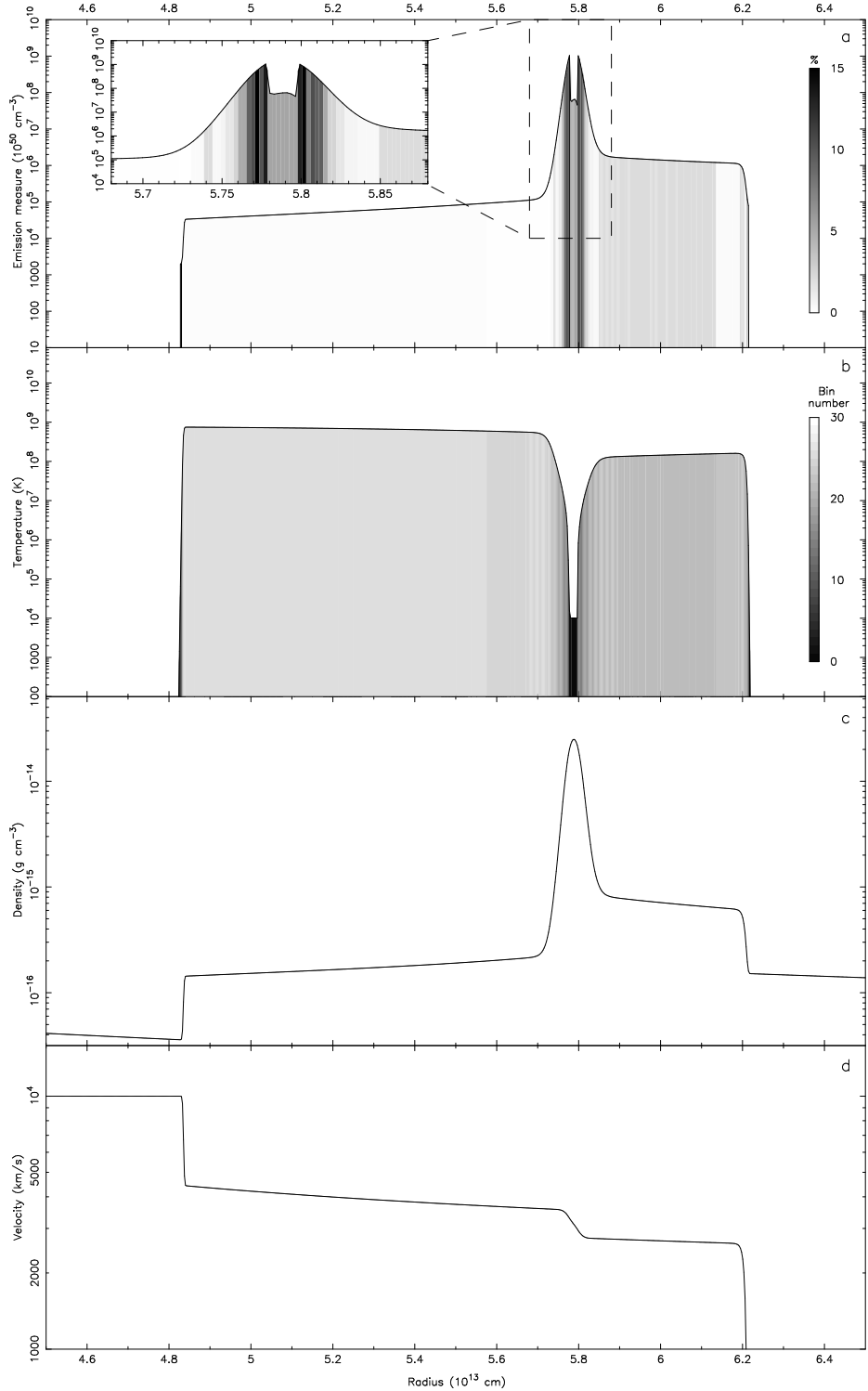}
\caption[Run 81 emission measures]{(a) Emission measure as a function of radius inside the hot shell for run 81 at day 3 after outburst. The data are greyscale coded to represent what percentage of $\text{EM}_{\text{tot}}$ each temperature bin represents. The scaling is shown on the right (ranging $0-15\%$). The inset shows the detail of the region around the contact discontinuity, which provides the majority of $\text{EM}_{\text{tot}}$. The temperature bin in which the post-shock material lies contributes only a very small fraction to $\text{EM}_{\text{tot}}$ (less than 5\%). (b) Gas temperature as a function of radius. The data are greyscale coded to represent which of the 30 temperature bins each part of the hot shell belongs to. The lower panels show the gas density (c) and velocity (d) as a function of radius.}
\label{fig:em}
\end{center}
\end{figure*}

Figure~\ref{fig:em}a shows the emission measure as a function of radius inside the expanding remnant for run 81 at day 3 after outburst. As mentioned in section~\ref{sec:calcxrayspec}, to calculate our synthetic X-ray spectra we have run through the radial cells of our grid binning the emission measures into 30 temperature bins evenly (logarithmically) spaced between $10^{4}-10^{9}$ K. The greyscale in Fig.~\ref{fig:em}a represents the percentage of the total emission measure contained within each temperature bin. The lower panels show temperature (b), density (c) and velocity (d) of the gas. The forward and reverse shocks are clearly visible at $6.22 \times 10^{13}$~cm and $4.83 \times 10^{13}$~cm, respectively. The contact discontinuity between the shocked ejecta and shocked RG wind (high density, well cooled and therefore low temperature) is located at $5.8 \times 10^{13}$ cm.

The major contribution to the total emission measure is from the region around the contact discontinuity, while the broader forward shocked region with temperatures of $\sim 10^{8}$ K (spanning radii from $5.9 - 6.2 \times 10^{13}$ cm) contributes only a few percent. \citet{orlando09} also find in their 3D simulations that most of the X-ray emission originates from a small dense region behind the forward shock. This implies that the total X-ray emission is dominated by material at lower temperatures (around the contact discontinuity) and that the contribution from the hotter forward shocked gas is small. However, the shocked gas might still contribute strongly to the high energy X-rays.

The dot-dash line in the top left panel of Fig.~\ref{fig:spec81} shows the X-ray spectrum emitted by the  gas located between the contact discontinuity and the forward shock (we shall call this the post-forward shock gas) with no circumstellar absorption included; the contribution to the total X-ray spectrum is small at low energies but is of the order of 50\% for energies above 5 keV. Such a contribution is non-negligible and should have a large impact on the temperature fits to the data. However, the temperature of the emitting gas determines the shape of an X-ray spectrum while the emission measure and distance to the source govern the normalisation of the spectrum.

The post-forward shock gas spectrum is much flatter than the total X-ray spectrum (solid line); by fitting a single temperature MEKAL model to the post-shock gas spectrum at day 3 we obtain $k_{B}T= 13.5$ keV, translating as a shock velocity of 3400\kms which is entirely consistent with the real shock velocity measured in the simulation (black dots in Fig.~\ref{fig:shockvel}a). This flat post-forward shock spectrum gets drowned by the bright but cooler emission from the rest of the shell which changes the shape of the X-ray spectrum and therefore changes the fit temperature.

Fitting temperatures to the total observed spectra in order to estimate post-shock temperatures and therefore shock velocities is thus unreliable. It also suggests that cosmic ray acceleration at the shock may not be required to explain the discrepancies between velocities derived from observed IR line widths and those derived from X-ray temperature fits (even though such acceleration might still take place).

Finally, the velocity plot in Fig.~\ref{fig:em}d shows that the material moving at the fast ejection speed of $10^{4}\kms$ contributes nothing to the X-ray emission and will thus not be observed, while the post reverse shock gas does emit a small amount and its velocities are similar to the ones measured by the IR line widths (as shown in Fig.~\ref{fig:shockvel}a).

\subsection{The RXTE dataset}\label{sec:rxtedata}

It has to be noted that even though this study primarily addresses the {\it Swift} X-ray data, one cannot ignore the higher energy X-ray data obtained with the RXTE observatory \citep{sokoloski06}. There are indeed some differences between the two datasets, taken during the same epoch of the remnant's evolution. The break in the {\it Swift} data around day 7-8 which we interpreted as the end of our fast wind phase \citepalias{vaytet07} is absent from the RXTE data. Instead, a nearly perfect Sedov-Taylor shock deceleration is observed between days 3 and 10 (see Fig.~2b in \citealt{sokoloski06}; note that once again these velocities were computed from single temperature fits to X-ray spectra). The discrepancy between the data is not fully understood. The RXTE probes a higher energy range ($3-20$ keV) than the XRT onboard {\it Swift} ($0.3-10$ keV) which would imply that it is sensitive to only the hottest plasma and would therefore give a better estimate of the post-shock temperatures and hence the shock velocities. However, the observed velocities are lower than the velocities derived from {\it Swift} observations \citepalias{bode06} which is inconsistent with our previous analysis showing that the single temperature fits to the {\it Swift} data underestimate the postshock temperatures in the system.

Another important difference between the datasets is that while an increase in total integrated flux is reported for the {\it Swift} data between days 3 and 5, a decrease is observed by the RXTE. This can be explained by the fact that a large portion of the energy radiated at day 3 is in the high energy range (thus not well detected by {\it Swift}) and then as the blastwave cools adiabatically the radiation moves to lower energies resulting at day 5 in an increase of $5-10$ keV photon counts for {\it Swift} and a decrease of $7-20$ keV photon counts for RXTE. In addition, as the forward shock progresses outward clearing the RG wind, the amount of circumstellar absorption reduces which also contributes to an increase of soft X-ray flux. An increase of $3-5$ keV photon counts is observed in the RXTE data which confirms this statement. However, if one compares the same energy band in the two datasets between days 3 and 5, one can see that the X-ray counts between $5-10$ keV increases in the {\it Swift} data while it decreases in the RXTE data. Inconsistent datasets will inevitably be difficult to fit with a single model.

The reason why a Sedov-Taylor model fits the early RXTE data and not the {\it Swift} is unclear. A general Sedov-Taylor blastwave is produced by an instantaneous release of energy at a point inside an ambient medium with a power-law density profile. No mass is ejected, there is no reverse shock, and the expanding shell is made up only from the ambient gas swept up by the shock. In our model, we have adopted a more physically realistic scenario likely to be applicable to the case of RS Oph where some mass has been ejected in the outburst. Here, the outburst energy is  delivered via the kinetic energy of a wind which then interacts with a circumstellar medium with a density described by an inverse-square law as produced by a stellar wind. We have experimented with varying the amounts of outburst energy in kinetic energy or in thermal energy. For example, we have run a simulation using the parameters from run 2 (same $M_{\mathrm{ej}}$ and $E_{0}$) but with half of the outburst energy in thermal form and half in kinetic form. The results are quite different at early times, while remaining very similar at later times ($t > 30$ days). At early times, the blastwave expands much faster (due to the instantaneous release of half the outburst energy), the X-ray spectra are much flatter (yielding higher X-ray temperatures) and the velocities are much more Sedov-like. As opposed to the initial run 2, there is a decrease of X-ray flux between days 3 and 5 when folded through both the {\it Swift} instrument response and the RXTE response. This is more in line with the RXTE observations but no longer in agreement with the {\it Swift} data. A more extreme example injecting 100\% of the outburst energy in thermal form (as in the models of \citealt{obrien92}) also shows a decrease in $5-10$ keV photon counts when folded through the {\it Swift} response. We have not been able to obtain results which are consistent with both the {\it Swift} and RXTE observations.

\section{Conclusions}\label{sec:conclusion}

We have taken the results from our hydrodynamical models described in \citetalias{vaytet07} and calculated synthetic X-ray spectra using the \textsc{xspec}. The results from runs 1 to 16 of \citetalias{vaytet07} were first compared to the {\it Swift} spectra. It was found that the rate of decrease of the hard X-ray counts was related to the rate of decrease of the forward shock velocities, and a best fitting model would need a relatively low ejecta mass. However, runs with very high ejection velocities $V_{2}$ overestimated the X-ray counts across the spectrum at all times. Run 13 which was thought to be the best fit (based on the forward shock velocities) in \citetalias{vaytet07} was thus excluded.

A wider exploration of the 1D model parameter space was undertaken to find a model which best fit the observed spectra. A total of 88 runs were carried out, and fluxes were extracted for each day between day 1 and day 100 after outburst for each run. These model fluxes were compared to the fluxes from 70 {\it Swift} observational epochs during the same period, being careful to exclude SSS emission as appropriate. Large differences between the observed and theoretical fluxes allowed us to exclude many runs, and the spectra of the few remaining runs with good matching fluxes were examined in order to find a best-fitting model. We are now confident that the mass ejected in the RS Oph outburst is relatively low, around $(2-5) \times 10^{-7}~\msun$, and that the shock velocities are higher than suggested by X-ray single temperature fits. It was however never possible to simultaneously obtain a sufficiently high level of hard X-ray emission, a flat mid-energy spectrum and an appropriately low level of emission at low energies without including some additional ad-hoc absorption, thus making the derived densities in the RG wind uncertain.

The fast wind velocity of the best-fitting model ($10000\kms$) is considered unusually high for a nova outburst, but radio observations of \citet{obrien06} and \citet{sokoloski08} show evidence for bipolar lobes moving at speeds of the order of $5000\kms$ at early times ($t < 50$ days). Chandra observations of \citet{luna09} also report a jet moving at speeds greater than $6000\kms$. Similarly, \citet{ribeiro09} deduce high ejection velocities of $\sim 5100\kms$ from HST imaging and ground-based optical spectra, albeit the latter two studies are based on observations taken much later (538 and 155 days after outburst, respectively) than the timescales addressed in this paper. Furthermore, the fast ejection velocity may not be easily directly observed as the ejecta are decelerated when they run into the circumstellar medium.

Another important conclusion we can draw from this study is that in the Swift/XRT energy band, the emission from the post-shock material is negligible compared to that of the regions around the high-density contact discontinuity, and that fitting emission models with single temperatures (even when several components with different temperatures are included) to X-ray spectra greatly underestimates immediate post-shock temperatures and hence forward shock velocities. We also showed that velocity estimates from line widths (for example the IR line widths in the case of RS Oph) also underestimate the forward shock velocities.

Some inconsistencies are however present between the {\it Swift} and RXTE X-ray datasets, in particular in the forward shock velocities between days 3 and 10. While the spectra of our best fitting model agree well with the {\it Swift} observations, they do not fit the RXTE data. This does not affect our results for the subsequent evolution of the remnant (in particular the ejected mass) since RS Oph was very faint in the RXTE energy range after day 10, as opposed to {\it Swift}. It just means that the manner in which the mass and energy are ejected during the outburst might be different (for example, the proportions of outburst energy in the form of kinetic and thermal energy may be different).

It has proved difficult to find a set of parameter values for this spherically symmetric  model which matches the observed X-ray spectra for all epochs during the first 100 days of the remnant's evolution. However, radio and optical observations have shown that the remnant of RS Oph has a bipolar morphology. Our suggestion that the additional absorption required to fit the low-energy X-ray spectra might be present in the form of a torus would also fit with an evolving bipolar morphology. Preliminary work on 2D codes with a bipolar ejection is underway, using an equatorial density enhancement in the RG wind. Initial results have shown that is it possible to obtain fast moving material in the polar direction yielding hard X-rays and a high absorption column along the equator at the same time, better resembling the observed X-ray spectra of RS Oph \citep{vaytet09}. A bipolar model might also be able to fit the RXTE observations better if the remnant exhibits more of a Sedov-Taylor blastwave behaviour along one direction. Much work however remains, due to the increasingly large parameter space worthy of exploration.

\acknowledgments

The authors are very grateful to the Swift Mission Operations Center staff for their superb efforts in supporting the observations reported here. NMHV greatfully acknowledges support from research grant ANR-06-CIS6-009-01 for the SiNeRGHy project at the CEA/SAp Saclay. The authors would also like to thank the referee J. Sokoloski for very useful comments which have helped to improve this paper and providing the data on the size of the expanding ring from the radio observations in Fig.~\ref{fig:shockpos}.


\begin{thebibliography}{}
\bibitem[\protect\citeauthoryear{Abbey et~al.}{2006}]{abbey06} Abbey, A., et~al. 2006, in Proc. The X-Ray Universe 2005 (ESA SP-604; Fekel, F.~C., Joyce, R.~R., Hinkle, K.~H., \& Skrutskie, M.~F. 2000, AJ, 119, 1375 Noordwijk: ESA), 943

\bibitem[\protect\citeauthoryear{Anupama}{2008}]{anupama08} Anupama, G.~C. 2008, in RS Ophiuchi (2006) and the Recurrent Nova Phenomenon, eds Evans A., Bode M.~F., O'Brien T.~J., Darnley M.~J., ASP Conference Series, 401, 31

\bibitem[\protect\citeauthoryear{Balman, Krautter \& Oegelman}{1998}]{balman98} Balman, S., Krautter, J., Oegelman, H. 1998, ApJ, 499, 395

\bibitem[\protect\citeauthoryear{Barry et~al.}{2008}]{barry08} Barry, R.~K., Mukai, K., Sokoloski, J.~L., Danchi, W.~C., Hachisu, I., Evans, A., Gehrz, R., Mikolajewska, J. 2008, in RS Ophiuchi (2006) and the Recurrent Nova Phenomenon, eds Evans A., Bode M.~F., O'Brien T.~J. \& Darnley M.~J., ASP Conference Series, 401, 52

\bibitem[\protect\citeauthoryear{Bode}{1987}]{bode87} Bode, M.~F. 1987, in RS Oph (1985) and the Recurrent Nova Phenomenon, ed. M.~F. Bode (Utrecht: VNU Science), 241

\bibitem[\protect\citeauthoryear{Bode et~al.}{2006}]{bode06} Bode, M.~F. et~al. 2006, ApJ, 652, 629 (Paper I)

\bibitem[\protect\citeauthoryear{Bode et~al.}{2007}]{bode07} Bode, M.~F., Harman, D.~J., O'Brien, T. J., Bond, H.~M., Starrfield, S., Evans, A., Eyres, S.~P.~S. 2007, ApJL, 655, L63

\bibitem[\protect\citeauthoryear{Das, Banerjee \& Ashok}{Das et~al.}{2006}]{das06} Das, R., Banerjee, D.~P.~K., Ashok N.~M. 2006, ApJL, 653, L141

\bibitem[\protect\citeauthoryear{Dobrzycka \& Kenyon}{1994}]{dobrzycka94} Dobrzycka, D., Kenyon, S.~J. 1994, AJ, 108, 2259

\bibitem[\protect\citeauthoryear{Evans et~al.}{2007}]{evans07} Evans, A. et~al. 2007, ApJL, 671, L157

\bibitem[\protect\citeauthoryear{Fekel et~al.}{2000}]{fekel00} Fekel, F.~C., Joyce, R.~R., Hinkle, K.~H, Skrutskie, M.~F. 2000, AJ, 119, 1375

\bibitem[\protect\citeauthoryear{Hachisu et~al.}{2006}]{hachisu06} Hachisu, I. et~al. 2006, ApJL, 651, L141

\bibitem[\protect\citeauthoryear{Hachisu, Kato \& Luna}{Hachisu et~al.}{2007}]{hachisu07} Hachisu, I., Kato, M., Luna, G.~J.~M. 2007, ApJL, 659, L153

\bibitem[\protect\citeauthoryear{Hjellming et~al.}{1986}]{hjellming86} Hjellming, R.~M., van Gorkom, J.~H., Seaquist, E.~R., Taylor, A.~R., Padin, S., Davis, R.~J., Bode, M.~F. 1986, ApJL, 305, L71

\bibitem[\protect\citeauthoryear{Krautter et~al.}{1996}]{krautter96} Krautter, J., Oegelman, H., Starrfield, S., Wichmann, R., Pfeffermann, E. 1996, ApJ, 456, 788

\bibitem[\protect\citeauthoryear{Lloyd et~al.}{1992}]{lloyd92} Lloyd, H.~M., O'Brien, T.~J., Bode, M.~F., Predehl, P., Schmitt, J.~H.~M.~M., Truemper, J., Watson, M.~G., Pounds, K.~A. 1992, Nature, 356, 222

\bibitem[\protect\citeauthoryear{Luna et~al.}{2009}]{luna09} Luna, G.~J.~M., Montez, R., Sokoloski, J.~L., Mukai, K., Kastner, J.~H. 2009, ApJ, 707, 1168

\bibitem[\protect\citeauthoryear{Mewe, Lemen \& van~den~Oord}{Mewe et~al.}{1986}]{mewe86} Mewe, R., Lemen, J.~R., van~den~Oord, G.~H.~J. 1986, A\&AS, 65, 511

\bibitem[\protect\citeauthoryear{Mukai \& Ishida}{2001}]{mukai01} Mukai, K., Ishida, M. 2001, ApJ, 551, 1024

\bibitem[\protect\citeauthoryear{Narumi et~al.}{2006}]{narumi06} Narumi, H., Hirosawa, K., Kanai, K., Renz, W., Pereira, A., Nakano, S., Nakamura, Y., Pojmanski, G. 2006, IAUC 8671

\bibitem[\protect\citeauthoryear{Ness et~al.}{2009}]{ness09} Ness, J.-U., Drake, J.~J., Starrfield, S., Bode, M.~F., O'Brien, T.~J., Evans, A., Eyres, S.~P.~S., Helton, L.~A., Osborne, J.~P., Page, K.~L., Schneider, C., Woodward, C.~E. 2009, AJ, 137, 3414

\bibitem[\protect\citeauthoryear{O'Brien et~al.}{2008}]{obrien08} O'Brien, T.~J., Beswick, R.~J., Bode, M.~F., Eyres, S.~P.~S., Muxlow, T.~W.~B., Garrington, S.~T., Porcas, R.~W., Evans, A., Davis, R.~J., 2008, in RS Ophiuchi (2006) and the Recurrent Nova Phenomenon, eds Evans A., Bode M.~F., O'Brien T.~J. \& Darnley M.~J., ASP Conference Series, 401, 239

\bibitem[\protect\citeauthoryear{O'Brien, Bode \& Kahn}{O'Brien et~al.}{1992}]{obrien92} O'Brien, T.~J., Bode, M.~F., Kahn, F.~D. 1992, MNRAS, 255, 683

\bibitem[\protect\citeauthoryear{O'Brien et~al.}{2006}]{obrien06} O'Brien, T.~J., Bode, M.~F., Porcas, R.~W., Muxlow, T.~W.~B., Eyres, S.~P.~S., Beswick, R.~J., Garrington, S.~T., Davis, R.~J., Evans, A. 2006, Nature, 442, 279

\bibitem[\protect\citeauthoryear{O'Brien, Lloyd \& Bode}{O'Brien et~al.}{1994}]{obrien94} O'Brien, T.~J., Lloyd, H.~M., Bode, M.~F. 1994, MNRAS, 271, 155

\bibitem[\protect\citeauthoryear{Oppenheimer \& Mattei}{1993}]{oppenheimer93} Oppenheimer, B.~D., Mattei, J.~A. 1993, JAAVSO, 22, 105

\bibitem[\protect\citeauthoryear{Orlando, Drake \& Laming}{Orlando et~al.}{2009}]{orlando09} Orlando, S., Drake, J.~J., Laming, J.~M. 2009, A\&A, 493, 1049

\bibitem[\protect\citeauthoryear{Osborne et~al.}{2006}]{osborne06} Osborne, J.~P. et~al. 2006, Astronomer's Telegram 770

\bibitem[\protect\citeauthoryear{Osborne et~al.}{2011}]{osborne11} Osborne, J.~P. et~al. 2008, ApJ, 727, 124

\bibitem[\protect\citeauthoryear{Page et~al.}{2008}]{page08} Page, K.~L., Osborne, J.~P., Beardmore, A.~P., Goad, M.~R., Wynn, G.~A., Bode, M.~F., O'Brien, T.~J. 2008, in RS Ophiuchi (2006) and the Recurrent Nova Phenomenon, eds Evans A., Bode M.~F., O'Brien T.~J. \& Darnley M.~J., ASP Conference Series, 401, 283

\bibitem[\protect\citeauthoryear{Ribeiro et~al.}{2009}]{ribeiro09} Ribeiro, V.~A.~R.~M., et~al. 2009, ApJ, 703, 1955

\bibitem[\protect\citeauthoryear{Rosino}{1987}]{rosino87a} Rosino, L. 1987, in RS Oph (1985) and the Recurrent Nova Phenomenon, ed. M.~F. Bode (Utrecht: VNU Science), 1

\bibitem[\protect\citeauthoryear{Rosino \& Iijima}{1987}]{rosino87b} Rosino, L., Iijima, T. 1987, in RS Oph (1985) and the Recurrent Nova Phenomenon, ed. M.~F. Bode (Utrecht: VNU Science), 27

\bibitem[\protect\citeauthoryear{Rupen, Mioduszewski \& Sokoloski}{Rupen et~al.}{2008}]{rupen08} Rupen, M.~P., Mioduszewski, A.~J., Sokoloski, J.~L. 2008, ApJ, 688, 559

\bibitem[\protect\citeauthoryear{Schaefer}{2004}]{schaefer04} Schaefer, B. 2004, IAUC, 8396

\bibitem[\protect\citeauthoryear{Schaefer}{2009}]{schaefer09} Schaefer, B. 2009, ApJ, 697, 721

\bibitem[\protect\citeauthoryear{Sokoloski et~al.}{2006}]{sokoloski06} Sokoloski, J.~L., Luna, G.~J.~M., Mukai, K., Kenyon, S.~J. 2006, Nature, 442, 276

\bibitem[\protect\citeauthoryear{Sokoloski et~al.}{2008}]{sokoloski08} Sokoloski, J.~L., Rupen, M.~P., Mioduszewski, ApJL, 2008, 685, L137

\bibitem[\protect\citeauthoryear{Starrfield}{2008}]{starrfield08} Starrfield, S. 2008, in Classical Novae, eds Bode M.~F. \& Evans A., Wiley, Cambridge University Press, Second edition, Chapt. 4

\bibitem[\protect\citeauthoryear{Tatischeff \& Hernanz}{2008}]{tatischeff08} Tatischeff, V., Hernanz, M. 2008, in RS Ophiuchi (2006) and the Recurrent Nova Phenomenon, eds Evans A., Bode M.~F., O'Brien T.~J., Darnley M.~J., ASP Conference Series, 401, 332

\bibitem[\protect\citeauthoryear{Vaytet}{2009}]{vaytet09} Vaytet, N.~M.~H. 2009, PhD Thesis, The University of Manchester

\bibitem[\protect\citeauthoryear{Vaytet, O'Brien \& Bode}{Vaytet et~al.}{2007}]{vaytet07} Vaytet, N.~M.~H., O'Brien, T.~J., Bode, M.~F., 2007, ApJ, 665, 654 (Paper II)

\bibitem[\protect\citeauthoryear{Walder, Folini \& Shore}{Walder et~al.}{2008}]{walder08} Walder, R., Folini, D., Shore, S.~N. 2008, A\&AL, 484, L9 

\bibitem[\protect\citeauthoryear{Wilms, Allen \& McCray}{Wilms et~al.}{2000}]{wilms00} Wilms, J., Allen, A., McCray, R., 2000, ApJ, 542, 914

\bibitem[\protect\citeauthoryear{Yaron et~al.}{2005}]{yaron05} Yaron, O., Prialnik, D., Shara, M.~M., Kovetz, A., 2005, ApJ, 623, 398

\end{thebibliography}
\end{document}